\documentclass[aps,prl,twocolumn,superscriptaddress,showpacs]{revtex4}
\usepackage{amsfonts}
\usepackage{amssymb}
\usepackage{amsmath}
\usepackage{graphicx}% Include figure files
\usepackage{dcolumn}% Align table columns on decimal point
\usepackage{bm}% bold math
\usepackage{flafter}

\begin{document}

\title{Dimerization-Induced Fermi-Surface Reconstruction in IrTe$_2$}

\author{Man Jin Eom}
\affiliation{Department of Physics, Pohang University of Science and Technology, Pohang 790-784, Korea}
\author{Kyoo Kim}
\email{kyoo@postech.ac.kr}
\affiliation{Department of Physics, Pohang University of Science and Technology, Pohang 790-784, Korea}
\author{Y. J. Jo}
\affiliation{Department of Physics, Kyungpook National University, Daegu 702-701, Korea}
\author{J. J. Yang}
\affiliation{Laboratory for Pohang Emergent Materials, Pohang University of Science and Technology, Pohang 790-784, Korea}
\affiliation{Max Plank POSTECH Center for Complex Phase Materials, Pohang University of Science and Technology, Pohang 790-784, Korea}
\author{E. S. Choi}
\affiliation{National High Magnetic Field Laboratory, Florida State University, Tallahassee, Florida 32310, USA}
\author{B. I. Min}
\affiliation{Department of Physics, Pohang University of Science and Technology, Pohang 790-784, Korea}
\author{J. -H. Park}
\affiliation{Department of Physics, Pohang University of Science and Technology, Pohang 790-784, Korea}
\affiliation{Max Plank POSTECH Center for Complex Phase Materials, Pohang University of Science and Technology, Pohang 790-784, Korea}
\author{S. -W. Cheong}
\affiliation{Laboratory for Pohang Emergent Materials, Pohang University of Science and Technology, Pohang 790-784, Korea}
\affiliation{Rutgers Center for Emergent Materials and Department of Physics and Astronomy, Piscataway, New Jersey 08854, USA}
\author{Jun Sung Kim}
\email{js.kim@postech.ac.kr}
\affiliation{Department of Physics, Pohang University of Science and Technology, Pohang 790-784, Korea}

\date{\today}

\begin{abstract}
We report a de Haas-van Alphen (dHvA) oscillation study on IrTe$_2$ single crystals showing complex dimer formations.
By comparing the angle dependence of dHvA oscillations with band structure calculations, we show distinct Fermi surface reconstruction induced by a 1/5-type and a 1/8-type dimerizations. This verifies that an intriguing quasi-two-dimensional conducting plane across the layers is induced by dimerization in both cases. A phase transition to the 1/8 phase with higher dimer density reveals that local instabilities associated with intra- and interdimer couplings are the main driving force for complex dimer formations in IrTe$_2$.
\end{abstract}

\maketitle
%1 Introduction - dimer formation
Multiple orbital degeneracy and its coupling to charge, spin, and lattice degrees of freedom often lead to the intriguing electronic phases with dimerization~\cite{NaTiO2:Takeda:lineardimer,LiVS2:Katayama:trimer,AlV2O4:Horibe:heptamer,MgTi2O4:Schmidt:helicaldimer,CuIr2S4:Radaelli:octamer}. Various types of dimerization, ranging from a stripe-type dimer~\cite{NaTiO2:Takeda:lineardimer} to a complicated octamer~\cite{CuIr2S4:Radaelli:octamer,CuIr2S4:Kojima:SOC}, have been observed in the systems with $t_{2g}$ orbital degeneracy, which are often called as valence bond solids. For compounds containing a high-$Z$ element Ir, the spatially-extended orbitals of 5$d$ electrons and their strong spin-orbit coupling introduce additional complexity.
%For example, CuIr$_2$S$_4$ exhibits the metal-insulator transition originating from the lattice dimerization of Ir$^{+4}$ pairs~\cite{CuIr2S4:Radaelli:octamer}, whose ground state is the spin-orbital entangled state~\cite{CuIr2S4:Kojima:SOC}.
%In this respect, investigating the nature of complex dimerization in the Ir compounds is essential for understanding the orbital-driven instability in the high-$Z$ transition metal compounds.
%2 Introduction - IrTe2 & previous study
IrTe$_2$ is a recent candidate of such systems, which shows an intriguing dimer formation~\cite{IrTe2:Yang:pd,IrTe2:Pyon:syn,IrTe2:Fang:kineticreduction,IrTe2:qian:vanhove,IrTe2:Oh:polymerization,IrTe2:Oh:soliton,IrTe2:ootsuki:ARPES,IrTe2:ootsuki:ARPES2,IrTe2:Huibao:GSstructure,IrTe2:Pascut:dimerization,IrTe2:Tatsuya:swiching} in a simple layered structure consisting of edge-sharing IrTe$_6$ octahedra shown in Fig.~1.
%At room temperature, the valence of Ir is found to be +3.5 with partially-filled 5$d$ orbital states (5$d^{5.5}$).
A first-order resistive transition occurs at $T$$_{s1}$$\sim$ 280 K, due to Ir-Ir and Te-Te dimer formations with a modulation vector $\overrightarrow{q}$=(1/5,0,-1/5)~\cite{IrTe2:Yang:pd}.
Several mechanisms for the so-called 1/5 dimerization have been proposed~\cite{IrTe2:Yang:pd,IrTe2:Fang:kineticreduction,IrTe2:qian:vanhove,IrTe2:ootsuki:ARPES,IrTe2:ootsuki:ARPES2,IrTe2:Oh:polymerization}, but yet to be clarified.
%In order to understand the so-called 1/5 dimerization, several mechanisms, including the Fermi surface (FS) nesting~\cite{IrTe2:Yang:pd}, the kinetic energy reduction due to Te band splitting~\cite{IrTe2:Fang:kineticreduction}, van hove singularity at the Fermi level ($E_F$)~\cite{IrTe2:qian:vanhove}, and the depolymerization-polymerization of anionic Te bonds~\cite{IrTe2:Oh:polymerization}, have been proposed, but yet to be clarified.
In this respect, clarifying the nature of complex dimerization in IrTe$_2$ leads to a better understanding on orbital-driven instabilities in high-$Z$ transition metal compounds.
\begin{figure}
\includegraphics*[width=7.5cm, bb=44 350 480 775]{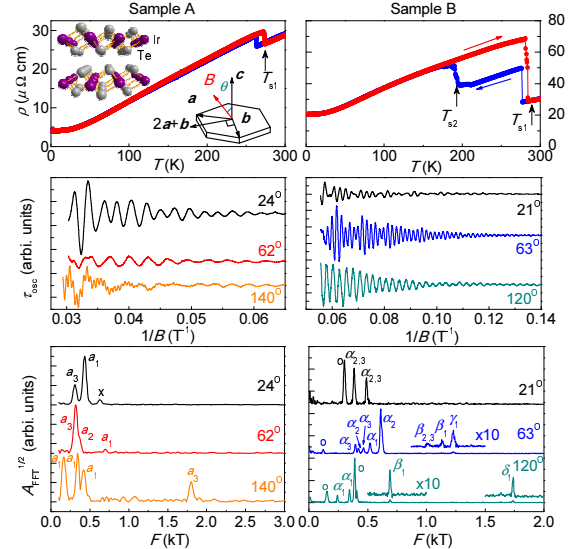}
\vspace{-3mm}
\caption{\label{fig1}(color online) The left and right columns are for the sample A and B, respectively. The top panels show the resistivity on cooling and warming. The transitions are indicated by the arrows. Crystal structure of IrTe$_2$ and the field orientation against the crystallographic axes are shown in the inset. The middle panels show the typical oscillatory torque signal ($\tau$$_{osc}$) for the sample A and B under different tilted $B$ up to 35 T and 18 T at $T$ = 0.5 K and 2 K, respectively. The bottom panels show the corresponding FFT peaks with labeling according to Figs.~2 and~3. The FFT data for the sample B are magnified in the high frequency region for clarity.}
\vspace{-3mm}
\end{figure}

\begin{figure*}
\includegraphics*[width=15cm, bb=5 595 595 800]{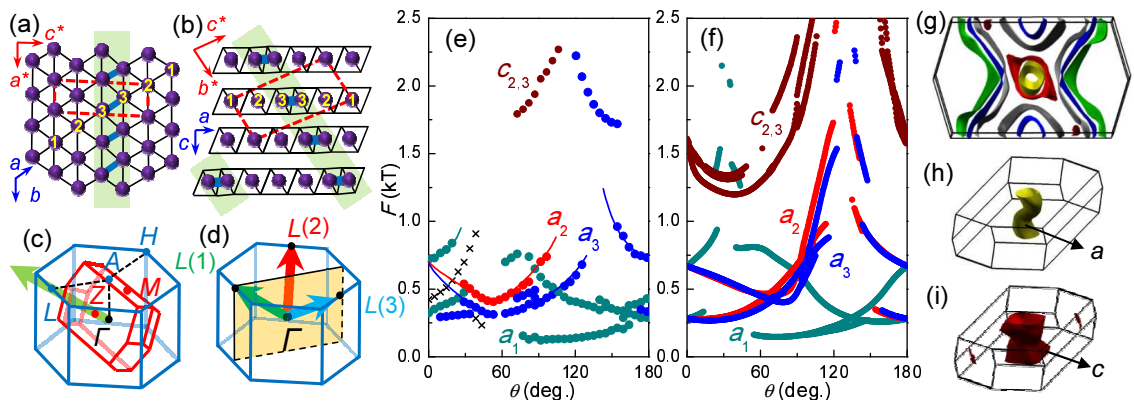}
\vspace{-1mm}
\caption{\label{fig2}(color online) Crystal structure of IrTe$_2$ (sample A) below the 1/5 dimerization at $T_{s1}$, showing (a) the triangular Ir layer and (b) the layer stacking order. The crystallographic axes for the high-$T$ ($a$, $b$, and $c$) and the low-$T$ ($a^*$, $b^*$, and $c^*$) structures are shown. The red box represents the unit cell of the 1/5 phase, and the numeric labels denote the inequivalent Ir sites. The Ir(3)-Ir(3) dimers and their stripes are marked with blue bonds and green shading, respectively. (c) The BZ of the high-$T$ (blue) and the low-$T$ (red) phases. The $\Gamma$-$Z$ symmetry line of the low-$T$ BZ is pointing to the direction of the $\Gamma$-$L$ line of the high-$T$ BZ. (d) Three possible directions of $\Gamma$-$Z$ lines of the low-$T$ BZ against the high-$T$ BZ (red, blue and green) with the corresponding domain index $n$ ($n$ = 1,2,3) in the parentheses. The shaded plane represents the basal plane of the rotating magnetic fields. The angle dependence of (e) the measured and (f) the calculated dHvA frequencies for the sample A. The cross symbols in (e) are unidentified frequencies. (g) The calculated FSs for the 1/5 phase. The letters in (e) and (f) refer to the Fermi pockets shown in (h) and (i), and their subscripts are the domain index $n$.}
\end{figure*}
Very recently, an additional phase transition in IrTe$_2$ was observed at $T$$_{s2}$ $\sim$180 K, well inside the 1/5 phase~\cite{IrTe2:Oh:soliton}. This suggests that the genuine ground state of IrTe$_2$ may have a different dimer pattern from the known 1/5 modulation. Identifying the dimer patterns and the corresponding electronic structures is crucial for clarifying the origin of complex dimer formation and the unusual physical properties in IrTe$_2$.
%3 Brief announcement - what we have done
%So far, several have been devoted to reveal the FS reconstruction for the 1/5 phase. However the domain formation after the structural transition hampers the detailed investigation on the electronic structures for the low-temperature phase.
In this Letter, we present the de Haas-van Alphen (dHvA) study on high-quality IrTe$_2$ single crystals. In contrast to angle-dependent photoemission spectroscopy~\cite{IrTe2:qian:vanhove,IrTe2:ootsuki:ARPES,IrTe2:ootsuki:ARPES2}, dHvA oscillations provide a direct access to the bulk Fermi surface (FS) of each domain after dimerization.
%We identify the FSs for two types of crystals showing different behaviors of resistive transitions; one shows only the 1/5 dimerization at $T_{s1}$$\sim$ 280 K, and the other shows additional transition at $T_{s2}$ $\sim$ 180 K.
By comparing the angle dependence of the dHvA oscillations with band structure calculations, we verified that the intriguing quasi two-dimensional (2D) state \emph{across the layers} is induced by the dimerization. Furthermore, for the low temperature phase below $T_{s2}$, we found that the dimer pattern with a modulation vector $\overrightarrow{q}$= (1/8, 0, -1/8) can explain the observed dHvA results. The additional transition to the 1/8 phase with higher dimer density reveals the local instability associated with intra- and inter-dimer coupling as the driving force for the transitions, rather than electronic instability near the Fermi level. Our findings strongly suggest that IrTe$_2$ is a rare example where the valence bond ordering and metallicity coexist.

%4 experimental methods
We investigated two types of IrTe$_2$ single crystals grown using the Te-flux method with different postcooling procedures, either quenching (Sample A) or slow cooling (Sample B)~\cite{supp}. The details on experiments and calculations are provided in the Supplemental Materials~\cite{supp}. As shown in Fig.~1, they show different behaviors of the resistive transitions. For the sample A, a single transition is clearly observed at $T$$_{s1}$$\sim$ 280 K, consistent with the previous reports~\cite{IrTe2:Oh:polymerization}. In contrast, the sample B exhibits two successive transitions at $T$$_{s1}$$\sim$ 280 K and $T_{s2}$ $\sim$ 180 K on cooling. On warming, unusually large thermal hysteresis is observed up to $T$$_{s1}$ $\sim$ 280 K.

For both samples we obtained clear dHvA oscillations. The typical oscillating torque signal ($\tau$$_{osc}$) and the corresponding fast Fourier transform (FFT) are shown in Fig.~1. The tilted angle of the magnetic field against the crystallographic axes of the high-$T$ phase is given in the inset of Fig. 1. Note that $\theta$ = 0$^{\circ}$ corresponds to $B$$\parallel$$c$, while $\theta$ = 90$^{\circ}$ corresponds to $B$$\perp$$c$ and $B$$\perp$$b$ ($i.e.$ $B$ $\parallel$ $2a+b$). The sample B shows much stronger dHvA oscillations, resulting narrower FFT peaks than the sample A~[Fig. 1]. The quantum scattering time for the FS with similar size of $F$ $\sim$0.4 kT is found to be a factor of two larger for the sample B than the sample A~\cite{supp}. This confirms that the slowly-cooled sample B is a better-quality crystal than the quenched sample A~\cite{res:note}. From the Onsager relation, $F$$=$($\Phi_0$/2$\pi^2$)$S_F$, where $\Phi_0$ is the flux quantum and $S_F$ is the cross-sectional area of the FS normal to the field, we found that the typical size of the FS is a few nm$^{-2}$, much smaller than the FS for the high-$T$
phase~\cite{IrTe2:Yang:pd,IrTe2:Fang:kineticreduction}. Our dHvA results, therefore, manifest dimerization-induced FS reconstruction in both cases.

We first discuss the sample A showing the transition at $T$$_{s1}$$\sim$ 280 K only. In this case, the 1/5 phase is stabilized at low temperature, as illustrated in Figs.~2(a) and (b)~\cite{IrTe2:Huibao:GSstructure,IrTe2:Pascut:dimerization,IrTe2:Tatsuya:swiching}.
In this phase, Ir-Ir dimerization at the Ir(3) sites occurs along the $a$-axis together with Te-Te dimerization (not shown), accompanied by charge transfer from Ir to Te atoms~\cite{IrTe2:ootsuki:ARPES,IrTe2:Pascut:dimerization}. The Ir ions with relatively larger valence (Ir$^{4+}$ like) form dimers running along the $b$-axis ($a^*$-axis), while those with smaller valence (Ir$^{3+}$ like) do not participate in dimerization. This leads to stripes of Ir atoms with different charges, and the sequence along the $a$-axis is "33344", where "3"("4") denotes the Ir$^{3+}$-like (Ir$^{4+}$-like) site~\cite{note:CO}. In the neighboring layer, the location of these stripes shifts by an unit cell, resulting a staircase-like arrangement with a modulation vector $\overrightarrow{q}=(1/5,0,-1/5)$. As a consequence, new Brillouin zone (BZ) after dimerization is tilted against the original BZ~[Fig.~2(c)]. The reconstructed FSs, therefore, are expected to be formed in a tilted BZ, which can be mapped out from the angle dependent dHvA frequencies.

\begin{figure*}
\includegraphics*[width=15cm, bb=5 559 595 812]{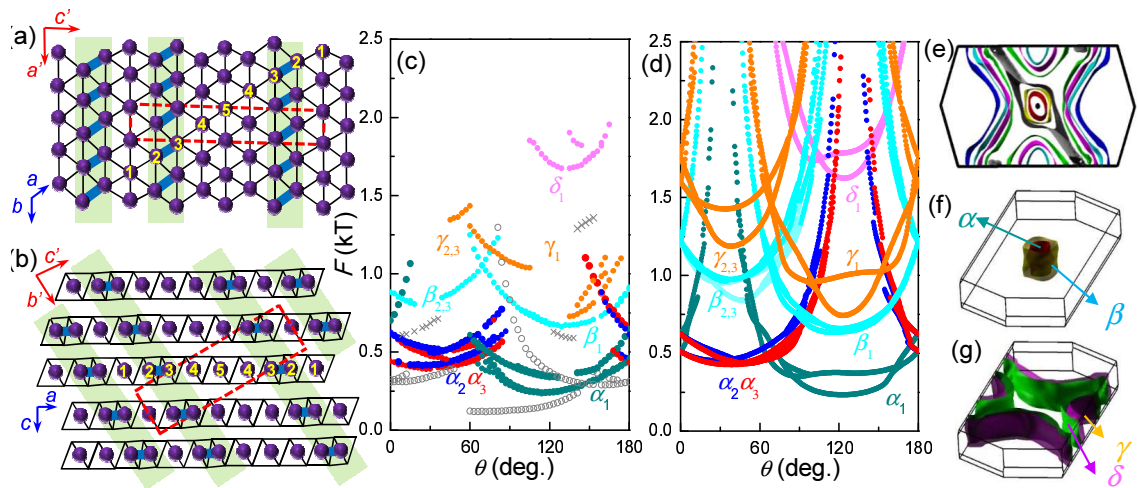}
\vspace{-1mm}
\caption{\label{fig3}(color online) The crystal structure of IrTe$_2$ (sample B) below the 1/8 dimerization at $T_{s2}$, showing (a) the triangular Ir layer and (b) the layers stacking. The crystallographic axes for the high-$T$ ($a$, $b$, and $c$) and the low-$T$ ($a'$, $b'$, and $c'$) structures are shown. The red box represents the unit cell of the 1/8 phase, and the numeric labels denote the inequivalent Ir sites. The Ir(2)-Ir(3) dimers and their stripes are marked with blue bonds and the green shading, respectively. The angle dependence of (c) the measured and (d) the calculated dHvA frequencies for the sample B. The open symbols in (c) are the dHvA frequencies found in the 1/5 phase, and the cross symbols are unidentified frequencies. (g) The calculated FSs for the 1/8 phase. The Greek letters in (c) and (d) refer to the Fermi pockets shown in (f) and (g), and their subscripts are the domain index $n$.}
\end{figure*}
By comparing the calculated and the observed dHvA frequencies for the 1/5 phase [Fig.~2(e) and (f)], we identify the corresponding FS pockets. Here, a magnetic field is rotated in the ($\Gamma$$LA$) plane of the original BZ, which means that the magnetic field is aligned once to the $\Gamma$-$Z$ direction of the new BZ in Fig.~2(c).
The low-frequency branches, denoted as $a_1$ [green in Figs.~2(e) and 2(f)], correspond to the orbits in the electron pocket centered at the $\Gamma$ point [Fig.~2(h)]. The larger electron pocket denoted as $c$ [Fig.~2(i)], is partially observed in Fig.~2(e). In addition, we found several dHvA frequencies [red and blue in Fig.~2(e)], whose angle dependences are not at all symmetric against the $c$-axis of the high-$T$ phase. These are due to two additional twin domains with relative 120$^{\circ}$ in-plane rotations. These are indicated by the domain index $n$ = 2 and 3, as compared to the former case with $n$ = 1. For these domains ($n$ = 2, 3), magnetic field is not aligned to any symmetry lines of the new BZ during rotation~[Fig.~2(d)]. Thus, the calculated dHvA frequencies for the $a$-FS from these domains, $a_2$ and $a_3$, show a distinct angle dependence from $a_1$ of the domain with $n$ = 1. Taking into account twin domains, we found good agreement between experiments and calculations as shown in Figs. 2(e) and 2(f). This suggests that the 1/5 phase has the quasi-2D FS as shown in Fig.2(g)~\cite{IrTe2:Pascut:dimerization,IrTe2:Tatsuya:swiching}.

The quasi-2D FSs in the tilted BZ implies that the 2D conducting plane is formed in the low-$T$ structure, which is declined by $\sim$ 10$^{\circ}$ with respect to the structural Ir and Te layers [Fig.~2(g)]. Such an unusual 2D system across the layers is due to the Ir-dimerization pattern~\cite{IrTe2:Pascut:dimerization,IrTe2:Tatsuya:swiching}. For the dimerized Ir sites, the local density of states (DOS) at $E_F$ is strongly suppressed, disconnecting the otherwise conducting layers. Since the stripes of the dimers have a staircase-like arrangement, as illustrated with the shaded regions in Fig.~2(b), the remaining Ir sites with relatively larger local DOS are connected across the layers and form a tilted 2D system. Our dHvA data, therefore, provide compelling experimental evidence for the intriguing 2D state across the layers in IrTe$_2$~\cite{IrTe2:Tatsuya:swiching}.

Now we discuss the sample B having two successive transitions at $T$$_{s1}$ $\sim$280 K and $T$$_{s2}$ $\sim$180 K. Figure 3(c) shows the angle dependence of dHvA frequencies for the sample B, which are much more complex than the case of the sample A. The same dHvA branches, found in the sample A, are also observed as indicated by grey symbols. However, several new dHvA branches, denoted as $\alpha$, $\beta$, $\gamma$ and $\delta$, are clearly observed. This implies that the electronic structure is reconstructed once again below $T$$_{s2}$ $\sim$180 K. The coexistence of the dHvA frequencies from the 1/5 phase and the new low-$T$ phase can be understood either by the magnetic breakdown effect~\cite{QO:shoenberg,note:MB} or by the phase separation~\cite{note:PS}. New dHvA frequencies for the sample B indicate that the low-$T$ phase below $T$$_{s2}$ $\sim$180 K has a distinct dimerization pattern from that of the 1/5 phase.
%Furthermore, considering that the magnetization is the thermodynamic quantity, this finding strongly suggests that the new low-$T$ phase in the sample B should be thermodynamically stable.

In order to understand the dHvA results below $T_{s2}$ for the sample B, we established a new structural model. Recently the scanning tunneling microscopy on the sample B showed that the 1/8-type dimer pattern is dominant over the 1/5-type below $T_{s2}$~\cite{IrTe2:Oh:soliton,IrTe2:Yeom}. The corresponding sequence of Ir dimers is therefore "34433344"-type, different from the "33344"-type sequence~\cite{IrTe2:Yeom} of the 1/5 phase. Taking this sequence into account, we impose the 1/8 modulation in the plane and the staircase-like arrangement along the out-of-plane direction as illustrated in Fig.~3(a) and (b). Recent single crystal X-ray scattering, in fact, observed the modulation vector $\overrightarrow{q}=(1/8,0,-1/8)$ for the sample B~\cite{IrTe2:JHPark}. Once the 1/8 structure is constructed by setting the Ir dimer patterns, internal parameters as well as total volume are relaxed in the calculations~\cite{supp}.

The resulting FSs with the 1/8 dimerization are shown in Fig.~3(e). As compared to the FSs of the 1/5 phase, more electron and hole pockets are newly formed, consistent with band folding effects with a large periodicity. Figure~3(d) shows the calculated dHvA frequencies for three 120$^{\circ}$-rotated domains as the case of the sample A. The $\alpha$$_n$ branches correspond to the smallest electron pocket centered at the $\Gamma$-point. The other electron pocket ($\beta$), slightly larger in size than the $\alpha$-pocket is also found in experiments. In addition, the two smallest hole pockets, $\gamma$ and $\delta$, centered at the $M$-point are partly observed. Good agreement between the measured and calculated dHvA frequencies as well as the cyclotron masses~\cite{supp}, especially for the small FSs, confirms that our structural model captures the correct electronic structures below $T_{s2}$. Similar to the 1/5 phase, the 1/8 phase also has quasi-2D FSs in a tilted BZ, hosting the intriguing cross-layer 2D conducting plane.

\begin{figure}
\includegraphics*[width=8cm, bb=30 430 390 780]{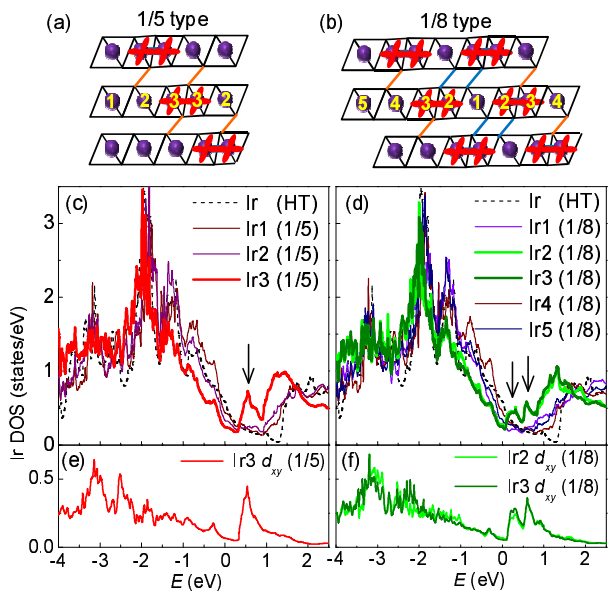}
\vspace{-1mm}
\caption{\label{fig4}(color online) The left and right columns are for the 1/5 and the 1/8 phase, respectively. The projection of (a) the 1/5 and (b) the 1/8 structures along the layer direction. Numeric labels denote the inequivalent Ir sites. The covalent bond is illustrated with the $d_{xy}$ orbitals in the Ir dimer. The orange and blue lines indicate the shortened Te-Te bonds. Partial density of states (DOS) of the Ir 5$d$ orbitals at the Ir sites in the low-$T$ phases, (c) the 1/5 and (d) the 1/8 phases. For comparison, we plot the Ir DOS of the high-$T$ phase (HT) together. The arrows indicate DOS peaks for the antibonding states of the Ir dimer. The DOS contribution of the 5$d_{xy}$ Ir orbitals involving dimerization are also shown for (e) the 1/5 and (f) the 1/8 phases.}
\end{figure}

%12 Figure.4 - Density of states
Having identified the FSs of the 1/5 and the 1/8 phases, we discuss the implication of our result to the origin of the dimerization in IrTe$_2$. First of all, our dHvA results show the 1/8 phase is stabilized in the better-quality crystal (sample B). Also, considering that magnetization is a thermodynamic quantity, the 1/8 phase identified by the dHvA effects should be thermodynamically stable. These results confirm that the 1/8 phase is not a meta-stable phase, but is rather a genuine ground state.  As illustrated in Fig.~3(b), the 1/8 phase has more Ir dimer stripes in the layers (two pairs from 8 Ir sites) than the 1/5 phase (one pair from 5 Ir sites)~\cite{note_jump}. These observations strongly suggest that the phase with a higher dimer density is favored at low temperature when the effects of impurities or local strains are minimized in IrTe$_2$~\cite{supp}.

Second, the additional transition to higher dimer density revealed that electronic instabilities near the $E_F$ are not responsible for the transition. For example, the instabilities due to FS nesting~\cite{IrTe2:Yang:pd} or a van-Hove singularity near the $E_F$~\cite{IrTe2:qian:vanhove} are relieved during the first transition to the 1/5 phase, which cannot induce the additional transition to the 1/8 phase. Then, the main driving force for dimerization should be the local instability related to the Ir or Te bonding. This contrasts to other $TmCh_2$ ($Tm$=transition metals, $Ch$= chalcogens), whose phase transition is mostly due to charge-density-wave formation. Rather, the low-$T$ phase in IrTe$_2$ is similar to the valence bond solids~\cite{NaTiO2:Takeda:lineardimer,LiVS2:Katayama:trimer,AlV2O4:Horibe:heptamer,MgTi2O4:Schmidt:helicaldimer,CuIr2S4:Radaelli:octamer}, where complex dimerization, associated with charge/orbital ordering, induces a "molecular" cluster state.

The band calculations further support our conclusions. As shown in Fig.~4, the 1/5 dimerization induces substantial change of the DOS in a wide energy range of $\pm$3 eV near $E_F$~\cite{IrTe2:Pascut:dimerization,IrTe2:Tatsuya:swiching}. In particular, there is a drastic bonding and antibonding splitting in the Ir(3)-Ir(3) dimer state, involving mostly $d_{xy}$ orbitals~[Fig.~4(c)], seen by a characteristic DOS peak of the antibonding state near $\sim$ 0.6 eV above the $E_F$~\cite{note_Te}. The bonding-antibonding splitting in the Ir dimer states is essential for lowering total energy~\cite{IrTe2:Pascut:dimerization,IrTe2:Tatsuya:swiching}, leading to Ir charge/orbital ordering in the 1/5 phase. For the 1/8 phase, two pairs out of 8 Ir ions form dimers, and they are close to each other as shown in Fig.~4(b). The characteristic antibonding DOS peak is further split by the 1/8 ordering~[Fig.~4(d)]. This arises from the coupling between the dimers~[Fig.~4(f)] through two additional shortened Te-Te bonds [the blue bonds in Fig.~4(b)] connecting two adjacent Ir(2)-Ir(3) dimers across the layer. Therefore, the inter-dimer coupling due to the spatially-extended Ir 5$d$ orbitals and their hybridization with Te orbitals drive the system towards the phase with higher dimer density. These results imply that the energy gain from the intra- and inter-dimer coupling compete with the energy loss from lattice distortion, and their subtle balance determines the phase transition in IrTe$_2$~\cite{note_phase}.

In conclusion, we identify the FSs for two types of the low-$T$ phases of IrTe$_2$, based on the dHvA oscillations and band structure calculations. For both cases of the 1/5 and the 1/8 modulations, the unusual quasi-2D conducting plane across the layers is induced by the staircase-like arrangement of the Ir and Te dimer stripes. We found that in a better-quality crystal, the additional phase transition to the 1/8 phase with a higher dimer density is induced. This clearly shows that the local instability for Ir charge and/or orbital ordering is the main driving force for the complex dimer formation in IrTe$_2$, similar to the valence bond solids. Our findings demonstrate that IrTe$_2$ provides a rare example where the valence bond ordering and metallicity coexist. How the coupling to itinerant electrons and the spin-orbit coupling play a role for stabilizing the 1/8 phase~\cite{supp} remains an important open question.

\emph{Note added.} Recently we became aware of the related dHvA study on IrTe$_2$ showing FS reconstruction~\cite{coldea}.

\begin{acknowledgments}
The authors thank H. W. Yeom and W. Kang for fruitful discussion. This work was supported by the NRF through the Mid-Career Researcher Program (Grant No. 2012-013838), SRC (Grant No. 2011-0030785), the Max Planck POSTECH/KOREA Research Initiative Program (Grant No. 2011-0031558), National Creative Initiative (Grant No. 2009-0081576), and also by IBS (No. IBS-R014-D1-2014-a02). B.I.M. and K.K. acknowledge support from the NRF Projects (Grant No. 2009-0079947 and No. 2011-0025237). SWC was supported by the NSF under Grant No. NSF-DMREF-1233349. A portion of this work was performed at the National High Magnetic Field Laboratory, which is supported by National Science Foundation Cooperative Agreement No. DMR-1157490, the State of Florida, and the U.S. Department of Energy.
\end{acknowledgments}


\begin{thebibliography}{00}
\bibitem{NaTiO2:Takeda:lineardimer} K. Takeda, K. Miyake, K. Takeda, and K. Hirakawa: J. Phys. Soc. Jpn. {\bf 61}, 2156 (1992).
\bibitem{LiVS2:Katayama:trimer} N. Katayama, M. Uchida, D. Hashizume, S. Niitaka, J. Matsuno, D. Matsumura, Y. Nishihata, J. Mizuki, N. Takeshita, A. Gauzzi, M. Nohara, and H. Takagi: Phys. Rev. Lett. {\bf 103}, 146405 (2009).
\bibitem{AlV2O4:Horibe:heptamer} Y. Horibe, M. Shingu, K. Kurushima, H. Ishibashi, N. Ikeda, K. Kato, Y. Motome, N. Furukawa, S. Mori, and T. Katsufuji, Phys. Rev. Lett. {\bf 96} 086406 (2006).
\bibitem{MgTi2O4:Schmidt:helicaldimer} M. Schmidt, W. Ratcliff, P. G. Radaelli, K. Refson, N. M. Harrison, and S. W. Cheong: Phys. Rev. Lett. {\bf 92}, 056402 (2004).
\bibitem{CuIr2S4:Radaelli:octamer} P. G. Radaelli, Y. Horibe, M. J. Gutmann, H. Ishibashi, C. H. Chen, R. M. Ibberson, Y. Koyama, Y.-S. Hor, V. Kiryukhin, and S.-W. Cheong, Nature {\bf 416}, 155 (2002).
\bibitem{CuIr2S4:Kojima:SOC} K. M. Kojima, R. Kadono, M. Miyazaki, M. Hiraishi, I. Yamauchi, A. Koda, Y. Tsuchiya, H. S. Suzuki, and H. Kitazawa, Phys. Rev. Lett. {\bf 112}, 087203 (2014).
\bibitem{IrTe2:Yang:pd} J. J. Yang, Y. J. Choi, Y. S. Oh, A. Hogan, Y. Horibe, K. Kim, B. I. Min, and S-W. Cheong, Phys. Rev. Lett. {\bf 108}, 116402 (2012).
\bibitem{IrTe2:Pyon:syn} S. Pyon, K. Kodo, and M. Nohara, J. Phys. Soc. Jpn. {\bf 81}, 053701 (2012).
\bibitem{IrTe2:Fang:kineticreduction} A. F. Fang, G. Xu, T. Dong, P. Zheng and N. L. Wang, Sci. Rep. {\bf 3}, 1153 (2013).
\bibitem{IrTe2:qian:vanhove} T. Qian \emph{et al.}, arXiv. 1311.4946 (2013).
\bibitem{IrTe2:ootsuki:ARPES} D. Ootsuki, Y. Wakisaka, S. Pyon, K. Kudo, M. Nohara, M. Arita, H. Anzai, H. Namatame, M. Taniguchi, N. L. Saini, and T. Mizokawa, Phys. Rev. B {\bf 86}, 014519 (2012).
\bibitem{IrTe2:ootsuki:ARPES2} D. Ootsuki \emph{et al.}, J. Phys. Soc. Jpn. {\bf 82}, 093704 (2013).
\bibitem{IrTe2:Oh:polymerization} Y. S. Oh, J. J. Yang, Y. Horibe, and S.-W. Cheong, Phys. Rev. Lett. {\bf 110}, 127209 (2013).
\bibitem{IrTe2:Huibao:GSstructure} H. Cao, B. C. Chakoumakos, X. Chen, J. Yan, M. A. McGuire, H. Yang, R. Custelcean, H. Zhou, D. J. Singh, and D. Mandrus, Phys. Rev. B {\bf 88}, 115122 (2013).
\bibitem{IrTe2:Oh:soliton} P.-J. Hsu, T. Mauerer, M. Vogt, J. J. Yang, Y. S. Oh, S.-W. Cheong, M. Bode, and W. Wu, Phys. Rev. Lett. {\bf 111}, 266401 (2013).
\bibitem{IrTe2:Pascut:dimerization} G. L. Pascut \emph{et al.}, Phys. Rev. Lett. {\bf 112}, 086402 (2014).
\bibitem{IrTe2:Tatsuya:swiching} T. Toriyama, M. Kobori, T. Konishi, Y. Ohta, K. Sugimoto, J. Kim, A. Fujiwara, S. Pyon, K. Kudo, and M. Nohara, J. Phy. Soc. Jpn. {\bf 83}, 033701 (2014).
\bibitem{supp} See Supplemental Material [URL will be inserted by publisher] for additional information regarding crystal growth, Lifshitz-Kosevich analysis on the dHvA oscillations, and band calculations.
\bibitem{res:note} The sample B has a larger residual resistivity than the sample A as shown in Fig.~1. However, for IrTe$_2$ the resistivity is mostly determined by the mosaicity of stripe domains rather than the impurity scattering rate. In this regards, the quantum scattering time, which is sensitive to the scattering events for each domain, is a more reliable means to test the crystal quality than the residual resistivity.
\bibitem{note:CO} Unlike CuIr$_2$S$_4$, the charge ordering in IrTe$_2$ is not complete due to the strong hybridization with the Te bands maintaining the metallic nature.
\bibitem{QO:shoenberg} D. Shoenberg, \emph{Magnetic Oscillations in Metals} (Cambridge Univ. Press, Cambridge, 1984)
\bibitem{note:MB} At high magnetic fields the electrons have sufficient cyclotron energy, and they can tunnel from one part of the FS to another across the small momentum gap as found in various charge-density-wave compounds.
\bibitem{note:PS} A recent STM study reveals coexistence of the domains with different periodicity on the surface in Ref. ~[\onlinecite{IrTe2:Oh:soliton}]. If the phase separation occurs in the bulk as well, the remnant 1/5 phase may contribute to the dHvA oscillations even below $T$$_{s2}$.
\bibitem{IrTe2:Yeom} H. S. Kim, T.-H. Kim, J. J. Yang, S.-W. Cheong, and H. W. Yeom, Phys. Rev. B {\bf90}, 201103(R) (2014).
\bibitem{IrTe2:JHPark} K. T. Ko \emph{et al.}, \emph{unpublished.}
\bibitem{note_jump} This is also consistent with the resistivity jump at $T_{s2}$ in Fig.~1(b) since more dimer stripes in the 1/8 phase electrically cut the structural planes enhancing the resistivity compared to the 1/5 phase.
\bibitem{note_Te} The change in the Te states, mainly from the shortened Te-Te bonds nearby the Ir dimers, is found to be less significant~\cite{IrTe2:Pascut:dimerization,supp}.
\bibitem{note_phase} This is also consistent with the fact that presence of the 1/8 phase strongly depends on subtle differences in the growth conditions.
\bibitem{coldea} S. F. Blake \emph{et al.} arXiv:1412.4163. 



\end{thebibliography}
\end{document}